\documentclass{svproc}
\usepackage{bm}
\usepackage{amsmath}
\usepackage{amsfonts}
\usepackage{amssymb}
\usepackage{acro}
\usepackage{wrapfig}
\usepackage{upgreek}
\usepackage{siunitx}
\usepackage{verbatim}
\usepackage[
nonumberlist,nopostdot,style=index] 
{glossaries}
\newglossary*{symbol}{Symbols}
\usepackage{makeidx}  
\makeindex
\usepackage{hyperref}
\usepackage[nameinlink]{cleveref} 
\usepackage[linesnumbered,ruled,noend,vlined]{algorithm2e}
\usepackage{algorithmic}
\usepackage[c3]{optidef}                    
\usepackage{subcaption} 
\usepackage{tikz}
\usetikzlibrary{arrows, decorations.pathreplacing, shapes, arrows, automata, calc, fit, positioning, backgrounds, external, arrows.meta}
\usetikzlibrary{patterns}
\usepackage{tkz-berge}

        \crefname{assumption}{assumption}{assumptions}
        \Crefname{assumption}{Assumption}{Assumptions}
\usepackage{url}

\usepackage{fancyhdr}
\usepackage{lipsum} 
\fancypagestyle{firstpage}{
  \fancyhf{} 
  \fancyfoot[C]{
    \scriptsize
    {This extended abstract is based on the original paper published in 2024 European Control Conference (ECC). For full details, including complete methodology and experimental results, please refer to the published work, \href{https://doi.org/10.1109/ITSC57777.2023.10422078}{DOI: ITSC57777.2023.10422078}
  }
}}

\DeclareAcronym{cav}{
    short = CAV,
    long  = connected and automated vehicle,
}
\DeclareAcronym{cg}{
    short = CG,
    long  = center of gravity
}
\DeclareAcronym{cdmpc}{
    short = coop. DMPC,
    long  = cooperative distributed model predictive control
}
\DeclareAcronym{cmpc}{
    short = CMPC,
    long  = centralized model predictive control
}
\DeclareAcronym{cpm}{
    short = CPM,
    long  = Cyber-Physical Mobility
}
\DeclareAcronym{cpmlab}{
    short = CPM Lab,
    long  = Cyber-Physical Mobility Lab
}
\DeclareAcronym{ctg}{
    short = CTG,
    long  = cost to go
}
\DeclareAcronym{ctc}{
    short = CTC,
    long  = cost to come
}
\DeclareAcronym{dag}{
    short = DAG,
    long  = directed acyclic graph
}
\DeclareAcronym{dds}{
    short = DDS,
    long  = data distribution service
}
\DeclareAcronym{dmpc}{
    short = DMPC,
    long  = distributed model predictive control
}
\DeclareAcronym{dpc}{
    short = DPC,
    long  = distributed predictive control
}
\DeclareAcronym{fsa}{
    short = FSA,
    long  = finite state automaton,
    short-indefinite = an,
}
\DeclareAcronym{fca}{
    short = FCA,
    long = future collision assessment,
    short-indefinite = an,
}
\DeclareAcronym{ffo}{
    short = FFO,
    long  = first-fit ordering,
    short-indefinite = an,
}
\DeclareAcronym{fov}{
    short = FOV,
    long  = field of view,
    short-indefinite = an,
}
\DeclareAcronym{fpv}{
    short = FPV,
    long  = first-person view,
    short-indefinite = an,
}
\DeclareAcronym{hdv}{
    short = HDV,
    long  = human-driven vehicle,
    short-indefinite = an,
}
\DeclareAcronym{hil}{
    short = HiL,
    long  = hardware-in-the-loop,
}
\DeclareAcronym{hlc}{
    short = HLC,
    long  = high-level controller,
    short-indefinite = an,
}
\DeclareAcronym{ldo}{
    short = LDO,
    long  = largest degree ordering,
    short-indefinite = an,
}
\DeclareAcronym{llc}{
    short = LLC,
    long  = low-level controller,
    short-indefinite = an,
}
\DeclareAcronym{lwa}{
    short = LWA*,
    long  = lazy weighted A*,
    short-indefinite = an,
}
\DeclareAcronym{lsp}{
    short = LazySP,
    long  = lazy shortest path,
    short-indefinite = an,
}
\DeclareAcronym{lra}{
    short = LRA*,
    long  = lazy receding horizon A*,
    short-indefinite = an,
}
\DeclareAcronym{mas}{
    short = MAS,
    long  = multi-agent system,
    short-indefinite = an,
}
\DeclareAcronym{mcts}{
    short = MCTS,
    long  = Monte Carlo tree search,
    short-indefinite = an,
}
\DeclareAcronym{mip}{
    short = MIP,
    long  = mixed integer programming,
    short-indefinite = an,
}
\DeclareAcronym{mil}{
    short = MiL,
    long  = model-in-the-loop,
}
\DeclareAcronym{milp}{
    short = MILP,
    long  = mixed integer linear programming,
    short-indefinite = an,
}
\DeclareAcronym{mld}{
    short = MLD,
    long  = mixed logical dynamical,
    short-indefinite = an,
}
\DeclareAcronym{mlc}{
    short = MLC,
    long  = mid-level controller,
    short-indefinite = an,
}
\DeclareAcronym{mp}{
    short = MP,
    long  = motion primitive,
    short-indefinite = an,
}
\DeclareAcronym{mpa}{
    short = MPA,
    long  = motion primitive automaton,
    short-indefinite = an,
}
\DeclareAcronym{mpc}{
    short = MPC,
    long  = model predictive control,
    short-indefinite = an,
}
\DeclareAcronym{ncs}{
    short = NCS,
    long  = networked control system,
    short-indefinite = an,
}
\DeclareAcronym{nlp}{
    short = NLP,
    long  = nonlinear programming,
    short-indefinite = an,
}
\DeclareAcronym{ocp}{
    short = OCP,
    long  = optimal control problem,
    short-indefinite = an,
    long-indefinite = an,
}
\DeclareAcronym{odd}{
    short = ODD,
    long  = operational design domain,
    short-indefinite = an,
    long-indefinite = an,
}
\DeclareAcronym{ode}{
    short = ODE,
    long  = ordinary differential equation,
    short-indefinite = an,
    long-indefinite = an,
}
\DeclareAcronym{pdmpc}{
    short = \mbox{P-DMPC},
    long  = prioritized \acl{dmpc}
}
\DeclareAcronym{pil}{
    short = PiL,
    long  = processor-in-the-loop
}
\DeclareAcronym{qp}{
    short = QP,
    long  = quadratic programming,
}
\DeclareAcronym{rhgs}{
    short = RHGS,
    long  = receding horizon graph search,
    short-indefinite = an,
}
\DeclareAcronym{rhc}{
    short = RHC,
    long  = receding horizon control,
    short-indefinite = an,
}
\DeclareAcronym{rrt}{
    short = RRT,
    long  = rapidly-exploring random tree,
    short-indefinite = an,
}
\DeclareAcronym{rss}{
    short = RSS,
    long  = responsibility-sensitive safety,
    short-indefinite = an,
}
\DeclareAcronym{rti}{
    short = RTI,
    long  = real-time iteration,
    short-indefinite = an,
}
\DeclareAcronym{scdmpc}{
    short = SC-DMPC,
    long = Synchronization-Based Cooperative Distributed Model Predictive Control,
    short-indefinite = an
}
\DeclareAcronym{scp}{
    short = SCP,
    long  = sequential convex programming,
    short-indefinite = an,
}
\DeclareAcronym{scr}{
    short = SCR,
    long  = sequential convex restriction,
    short-indefinite = an,
}
\DeclareAcronym{sdo}{
    short = SDO,
    long  = saturation degree ordering,
    short-indefinite = an,
}
\DeclareAcronym{sgs}{
    short = SGS,
    long  = state-of-the-art graph search,
    short-indefinite = an,
}
\DeclareAcronym{sil}{
    short = SiL,
    long  = software-in-the-loop,
}
\DeclareAcronym{sl}{
    short = SL,
    long  = sequential linearization,
    short-indefinite = an,
}
\DeclareAcronym{sqp}{
    short = SQP,
    long  = sequential quadratic programming,
    short-indefinite = an,
}
\DeclareAcronym{tsp}{
    short = TSP,
    long  = traveling salesman problem,
}
\DeclareAcronym{uav}{
    short = UAV,
    long  = unmanned aerial vehicle,
    long-indefinite = an,
}
\DeclareAcronym{udlab}{
    short = IDS3C,
    long  = Information and Decision Science Scaled Smart City,
}
\DeclareAcronym{xil}{
    short = XiL,
    long  = X-in-the-loop,
    long-indefinite = an,
}
\DeclareAcronym{mycmpc}{
    short = CMPC,
    long  = Centralized Model Predictive Control
}
\DeclareAcronym{mycdmpc}{
    short = CDMPC,
    long  = Cooperative Distributed Model Predictive Control
}
\DeclareAcronym{myscdmpc}{
    short = SCDMPC,
    long  = Synchronization-Based Cooperative Distributed Model Predictive Control
}
\newcommand{\anAgent}{\ensuremath{i}}

\newcommand{\timestep}{\ensuremath{k}}
\newcommand{\timestepIterator}{\ensuremath{l}}

\newcommand{\ofAgent}[1]{\ensuremath{^{(#1)}}}

\newglossaryentry{matrix:Adjacency}{
	name=\ensuremath{\bm{D}},
	description={Adjacency matrix},
	sort={D},
    type=symbol
}

\newglossaryentry{set:realNumbers}{
	name=\ensuremath{\mathbb{R}},
	description={Set of real numbers},
	sort={real numbers},
    type=symbol
}

\newglossaryentry{set:naturalNumbers}{
	name=\ensuremath{\mathbb{N}},
	description={Set of natural numbers},
	sort={natural numbers},
    type=symbol
}

\newglossaryentry{set:systemStates}{
	name=\ensuremath{\mathcal{S}},
	description={Set of system states},
	sort={set of system states},
    type=symbol
}

\newglossaryentry{set:bigO}{
	name=\ensuremath{O},
	description={Big O},
	sort={O},
    type=symbol
}

\newglossaryentry{scalar:Weight}{
	name=\ensuremath{w},
	description={Weight},
	sort={weight},
    type=symbol
}
\newcommand{\weight}{\gls{scalar:Weight}}

\newglossaryentry{scalar:NumberOfAgents}{
    name=\ensuremath{N_A},
    description={Number of agents},
    sort={Number of agents},
    type=symbol
}

\newglossaryentry{graph:path}{
    name=\ensuremath{\pi},
    description={Path},
    sort={Path},
    type=symbol
}
\newcommand{\graphPath}{\gls{graph:path}}

\newglossaryentry{scalar:NumberOfVerticesInPath}{
    name=\ensuremath{N_{\graphPath}},
    description={Number of vertices in path $\graphPath$, or length},
    sort={Number of vertices in path},
    type=symbol
}

\newglossaryentry{trajectory:Reference}{
    name=\ensuremath{\bm{r}},
    description={Reference trajectory},
    sort={Reference Trajectory},
    type=symbol
}
\newcommand{\trajectoryReference}{\ensuremath{\gls{trajectory:Reference}}}

\newglossaryentry{sym:horizonControl}{
	name=\ensuremath{N_u},
	description={Control horizon in model predictive control},
	sort={Nu},
    type=symbol
}
\newcommand{\horizonControl}{\gls{sym:horizonControl}}

\newglossaryentry{sym:horizonPrediction}{
	name=\ensuremath{N_p},
	description={Prediction horizon in model predictive control},
	sort={Np},
    type=symbol
}
\newcommand{\horizonPrediction}{\gls{sym:horizonPrediction}}

\newglossaryentry{sym:vehicleOrientation}{
	name=\ensuremath{\psi},
	description={Vehicle orientation},
	sort={psi},
    type=symbol
}

\newglossaryentry{sym:sysModelContinuous}{
    name=\ensuremath{f},
    description={Continuous-time system model},
    sort={f continuous-time},
    type=symbol
}

\newglossaryentry{sym:sysModelDiscrete}{
    name=\ensuremath{f_{d}},
    description={Discrete-time system model},
    sort={f discrete-time},
    type=symbol
}

\newglossaryentry{sym:sysControlInputs}{
	name=\ensuremath{\bm{u}},
	description={System control inputs},
	sort=u,
    type=symbol
}
\NewDocumentCommand{\sysControlInputs}{ o }{\glslink{sym:sysControlInputs}{%
    \IfNoValueTF{#1}%
        {\ensuremath{\bm{u}}}%
        {\ensuremath{\bm{u}^{(#1)}}}%
}}

\newglossaryentry{sym:outputs}{
	name=\ensuremath{\bm{y}},
	description={System outputs},
	sort={y},
    type=symbol
}

\newglossaryentry{sym:sysSpeed}{
	name=\ensuremath{\mathrm{v}},
	description={Vehicle speed},
	sort={v},
    type=symbol
}
\newcommand{\sysSpeed}{\gls{sym:sysSpeed}}

\newglossaryentry{sym:inSpeed}{
	name=\ensuremath{u_{\sysSpeed}},
	description={Vehicle input speed},
	sort={uv},
    type=symbol
}

\newglossaryentry{sym:steering-angle}{
	name=\ensuremath{\delta},
	description={Vehicle steering angle},
	sort={delta},
    type=symbol
}

\newglossaryentry{sym:inSteering}{
	name=\ensuremath{u_{\delta}},
	description={Vehicle input steering angle},
	sort={ud},
    type=symbol
}

\newglossaryentry{sym:nColors}{
	name=\ensuremath{N_c},
	description={Number of colors},
	sort={Number of colors},
    type=symbol
}

\newglossaryentry{sym:nStates}{
	name=\ensuremath{n},
	description={Number of states of a dynamical system},
	sort={Number of states},
    type=symbol
}

\newglossaryentry{sym:nInputs}{
    name=\ensuremath{m},
    description={Number of inputs of a dynamical system},
    sort={m number of inputs},
    type=symbol
}

\newglossaryentry{sym:nLevels}{
	name=\ensuremath{N_{\text{CL}}},
	description={Number of computation levels},
	sort={Number of computation levels},
    type=symbol
}

\newglossaryentry{sym:nLevelsAllowed}{
	name=\ensuremath{N_{\text{CL,al.}}},
	description={Allowed number of computation levels},
	sort={Number of computation levels allowed},
    type=symbol
}

\newglossaryentry{sym:numGroups}{
	name=\ensuremath{N_{g}},
	description={Number of parallelly computing groups of agents},
	sort={Number of groups},
    type=symbol
}

\newglossaryentry{sym:fnPrio}{
    name=\ensuremath{p},
    description={Priority assignment function},
    sort={Priority assignment function},
    type=symbol
}

\newglossaryentry{sym:tSample}{
	name=\ensuremath{T_s},
	description={Sample Time},
	sort={T sample},
    type=symbol
}

\newglossaryentry{sym:tSolve}{
	name=\ensuremath{T_\text{sol.}},
	description={Computation time \tSolveB{\anAgent} that agent $\anAgent$ needs to solve its \ac{ocp}},
	sort={T solve},
    type=symbol
}

\newcommand{\tSolveB}[1]{\glslink{sym:tSolve}{\ensuremath{\ensuremath{T_\text{sol.}}^{(#1)}}}}

\newglossaryentry{sym:tSolveUpper}{
	name=\ensuremath{T_\text{sol.,max}},
	description={Upper computation time $T_\text{sol.,max}\ofAgent{\anAgent}$ that agent $\anAgent$ needs to solve it \ac{ocp}},
	sort={T solve upper},
    type=symbol
}

\newglossaryentry{sym:vertices}{
	name=\ensuremath{\mathcal{V}},
	description={Set of vertices},
	sort={Vertices},
    type=symbol
}
\newcommand{\setVertices}{\gls{sym:vertices}}
\newcommand{\setAgents}{\setVertices}

\newcommand{\helpSetPredecessors}[1]{\ensuremath{\setVertices^{(#1\leftarrow)}}}
\newglossaryentry{sym:predecessors}{
	name=\ensuremath{\helpSetPredecessors{i}},
	description={Set of predecessors of vertex $i$},
	sort={Vertices 1},
    type=symbol
}

\newcommand{\helpSetPredecessorsPar}[1]{\ensuremath{\setVertices^{(#1\leftarrow)}_{\text{par.}}}}
\newglossaryentry{sym:predecessorsPar}{
	name=\ensuremath{\helpSetPredecessorsPar{i}},
	description={Set of predecessors of vertex $i$ that have parallel couplings with it},
	sort={Vertices 2},
    type=symbol
}

\newcommand{\helpSetPredecessorsSeq}[1]{\ensuremath{\setVertices^{(#1\leftarrow)}_{\text{seq.}}}}
\newglossaryentry{sym:predecessorsSeq}{
	name=\ensuremath{\helpSetPredecessorsSeq{i}},
	description={Set of predecessors of vertex $i$ that have sequential couplings with it},
	sort={Vertices 3},
    type=symbol
}

\newcommand{\helpSetSuccessors}[1]{\ensuremath{\setVertices^{(#1\rightarrow)}}}
\newglossaryentry{sym:successors}{
	name=\ensuremath{\helpSetSuccessors{i}},
	description={Set of successors of vertex $i$},
	sort={Vertices 4},
    type=symbol
}

\newglossaryentry{sym:neighbors}{
	name=\ensuremath{\setVertices^{(i)}},
	description={Set of neighbors of vertex $i$},
	sort={Vertices 0},
    type=symbol
}

\newglossaryentry{sym:degree}{
	name=\ensuremath{d^{(i)}},
	description={Degree of vertex $i$. Sum of in-degree and out-degree},
	sort=degree,
    type=symbol
}

\newcommand{\helpVertexInDegree}[1]{\ensuremath{d^{(#1\leftarrow)}}}
\newglossaryentry{sym:inDegree}{
    name=\helpVertexInDegree{i},
    description={In-degree of vertex $i$},
    sort={degree in},
    type=symbol
}

\newcommand{\helpVertexOutDegree}[1]{\ensuremath{d^{(#1\rightarrow)}}}
\newglossaryentry{sym:outDegree}{
    name=\helpVertexOutDegree{i},
    description={Out-degree of vertex $i$},
    sort={degree out},
    type=symbol,
}

\newglossaryentry{sym:matLevels}{
	name=\ensuremath{\bm{L}},
	description={Matrix of computation levels},
	sort=L,
    type=symbol
}

\newglossaryentry{sym:tComp}{
	name=\ensuremath{T},
	description={Computation time},
	sort={T},
    type=symbol
}

\newglossaryentry{sym:tCompNcs}{
	name=\ensuremath{T_{\text{NCS}}},
	description={Computation time of \iac{ncs}},
	sort={T NCS},
    type=symbol
}

\newglossaryentry{graph:Undirected}{
	name=\ensuremath{\mathcal{G}},
	description={Undirected Graph},
	sort={graph1},
    type=symbol
}
\newcommand{\graphUndirected}{\gls{graph:Undirected}}

\newglossaryentry{graph:Directed}{
    name=\ensuremath{\vec{\gls*{graph:Undirected}}},
	description={Directed Graph},
	sort={graph2},
    type=symbol
}

\newglossaryentry{mat:edgeUtilities}{
    name=\ensuremath{M_\text{u}},
	description={Edge utility matrix},
	sort={matrix edge utilities},
    type=symbol
}

\newglossaryentry{sym:setColors}{
    name=\ensuremath{\mathcal{C}},
    description={Set of colors},
    sort=Colors,
    type=symbol
}

\newglossaryentry{sym:varControlInvariantSet}{
	name=\ensuremath{\mathcal{C}_{\text{inv}}},
	description={Control invariant set},
	sort={Control invariant set},
    type=symbol
}

\newglossaryentry{set:Weights}{
	name=\ensuremath{\mathcal{W}},
	description={Set of weights in a weighted graph},
    sort={Weights},
    type=symbol
}
\newcommand{\setWeights}{\gls{set:Weights}}

\newglossaryentry{set:Edges}{
	name=\ensuremath{\mathcal{E}},
	description={Set of edges; used to indicate that only undirected edges exist},
    sort={Edges},
    type=symbol
}
\newcommand{\setEdges}{\gls{set:Edges}}

\newglossaryentry{sym:setEdgesDirected}{
	name=\ensuremath{\vec{\gls*{set:Edges}}},
	description={Set of directed edges},
	sort={Edges directed},
    type=symbol
}

\newglossaryentry{sym:varEdge}{
	name=\ensuremath{(i \rightarrow j)},
	description={Directed edge from vertex $i$ to vertex $j$},
	sort={edge},
    type=symbol
}

\newglossaryentry{sym:fnReorder}{
	name=\ensuremath{f_r},
	description={Reordering function for graph color values},
	sort=fr,
    type=symbol
}

\newglossaryentry{sym:fcnObjective}{
    name=\ensuremath{J},
    description={Objective function of an optimization problem},
    sort=J,
    type=symbol
}
\NewDocumentCommand{\fcnObjective}{ o }{\glslink{sym:fcnObjective}{%
    \IfNoValueTF{#1}%
        {\ensuremath{J}}%
        {\ensuremath{J^{(#1)}}}%
}}

\newglossaryentry{sym:fcnObjectiveState}{
    name=\ensuremath{\ell_{x}},
    description={Reference tracking objective function},
    sort={lx Reference tracking objective function},
    type=symbol
}
\NewDocumentCommand{\fcnObjectiveState}{ o }{\glslink{sym:fcnObjectiveState}{%
    \IfNoValueTF{#1}%
        {\ensuremath{\ell_{x}}}%
        {\ensuremath{\ell_{x}^{(#1)}}}%
}}

\newglossaryentry{sym:fcnObjectiveStateTerminal}{
    name=\ensuremath{\ell_{x,f}},
    description={Reference tracking objective terminal function},
    sort={lf Reference tracking objective terminal function},
    type=symbol
}
\NewDocumentCommand{\fcnObjectiveStateTerminal}{ o }{\glslink{sym:fcnObjectiveStateTerminal}{%
    \IfNoValueTF{#1}%
        {\ensuremath{\ell_{x,f}}}%
        {\ensuremath{\ell_{x,f}^{(#1)}}}%
}}

\newglossaryentry{sym:fcnObjectiveInput}{
    name=\ensuremath{\ell_{u}},
    description={Input change objective function},
    sort={lu Input change objective function},
    type=symbol
}
\NewDocumentCommand{\fcnObjectiveInput}{ o }{\glslink{sym:fcnObjectiveInput}{%
    \IfNoValueTF{#1}%
        {\ensuremath{\ell_{u}}}%
        {\ensuremath{\ell_{u}^{(#1)}}}%
}}

\newglossaryentry{sym:fcnObjectiveCoupling}{
    name=\ensuremath{\ell_\text{c}},
    description={Coupling objective function},
    sort={lc Coupling objective function},
    type=symbol
}
\NewDocumentCommand{\fcnObjectiveCoupling}{ oo }{\glslink{sym:fcnObjectiveCoupling}{%
    \IfNoValueTF{#1}%
        {\ensuremath{\ell_\text{c}}}%
        {\ensuremath{\ell_\text{c}^{(#1,#2)}}}%
}}

\newglossaryentry{sym:fcnConstraintCoupling}{
    name=\ensuremath{c_\text{c}},
    description={Coupling constraint function},
    sort={cc Coupling constraint function},
    type=symbol
}
\NewDocumentCommand{\fcnConstraintCoupling}{ oo }{\glslink{sym:fcnConstraintCoupling}{%
    \IfNoValueTF{#1}%
        {\ensuremath{c_\text{c}}}%
        {\ensuremath{c_\text{c}^{(#1,#2)}}}%
}}

\newglossaryentry{sym:prediction}{
	name=\ensuremath{\tilde{\bm{x}}^{(j \leftarrow i)}_{\cdot \vert k}},
	description={Prediction in agent $i$ for agent $j$ at time $k$},
	sort=x,
    type=symbol,
}

\newglossaryentry{sym:state}{
	name=\ensuremath{\bm{x}},
	description={System state},
	sort=x,
    type=symbol
}
\newcommand{\sysState}{\gls{sym:state}}

\newglossaryentry{sym:stateAgent}{
	name=\ensuremath{\sysState^{(i)}_{(k)}},
	description={System state of agent $i$ at time $k$},
	sort=x,
    type=symbol,
}

\newglossaryentry{sym:ref}{
	name=\ensuremath{\sysState^{(i)}_{\text{ref},(k)}},
	description={System state reference of agent $i$ at time $k$},
	sort=x ref,
    type=symbol
}

\newglossaryentry{sym:setReachable}{
	name=\ensuremath{\mathcal{R}^{(i)}},
	description={reachable set of agent $i$},
	sort={Reachable set},
    type=symbol
}

\newglossaryentry{set:occupiedArea}{
	name=\ensuremath{\mathcal{O}^{(i)}},
	description={Set of the occupied area of the predicted trajectory of agent $\anAgent$},
	sort={occupied area},
    type=symbol
}

\newglossaryentry{set:feasibleStates}{
	name=\ensuremath{\mathcal{X}},
	description={set of feasible states},
	sort={x},
    type=symbol
}
\newcommand{\setFeasibleStates}{\gls{set:feasibleStates}}

\newglossaryentry{set:feasibleStatesTerminal}{
	name=\ensuremath{\mathcal{X}_f},
	description={set of feasible states at the prediction horizon},
	sort={x},
    type=symbol
}

\newglossaryentry{set:feasibleInputs}{
	name=\ensuremath{\mathcal{U}},
	description={set of feasible inputs},
	sort={u},
    type=symbol
}
\newcommand{\setFeasibleInputs}{\gls{set:feasibleInputs}}

\newglossaryentry{sym:numStatesConfSpace}{
    name=\ensuremath{n_p},
    description={Number of states that are in the conflictual space},
    sort={n number of states that are in the conflictual space},
    type=symbol
}

\newglossaryentry{sym:fnProj}{
    name=\text{proj},
    description={A function that projects a reachable set of system states in the conflictual space},
    sort={Project function},
    type=symbol
}

\begin{document}
\frontmatter          
\pagestyle{headings}  
%
%
\mainmatter              
\title{Synchronization-Based Cooperative Distributed Model Predictive Control}
\titlerunning{Synchronization-Based CDMPC}  
%
\author{Julius Beerwerth\inst{1}, Maximilian Kloock\inst{2} \and Bassam Alrifaee\inst{1}}
\authorrunning{Julius Beerwerth, Maximilian Kloock and Bassam Alrifaee} 
%
\tocauthor{Julius Beerwerth, and Bassam Alrifaee}
\index{Beerwerth, J.}
\index{Kloock, M.}
\index{Alrifaee, B.}
\institute{$^1$ Department of Aerospace Engineering,\\ University of the Bundeswehr Munich,\\ 85579 Neubiberg, Germany,\\
\email{[firstname].[lastname]@unibw.de}\\
$^2$ Department of Computer Science,\\ RWTH Aachen University,\\ 52062 Aachen, Germany\\}

\maketitle              
\thispagestyle{firstpage}

\begin{abstract}
Distributed control algorithms are known to reduce overall computation time compared to centralized control algorithms. However, they can result in inconsistent solutions leading to the violation of safety-critical constraints. 
Inconsistent solutions can arise when two or more agents compute concurrently while making predictions on each others control actions. To address this issue, we propose an iterative algorithm called Synchronization-Based Cooperative Distributed Model Predictive Control, which we presented in \cite{kloock_coordinated_2023}. The algorithm consists of two steps: 1. computing the optimal control inputs for each agent and 2. synchronizing the predicted states across all agents. We demonstrate the efficacy of our algorithm in the control of multiple small-scale vehicles in our Cyber-Physical Mobility Lab.
\end{abstract}
\section{Introduction}
    Distributed control algorithms in \ac{ncs} offer enhanced scalability, flexibility, and fault tolerance compared to \ac{mycmpc}. \ac{mycdmpc} is a popular approach, but it faces challenges with prediction inconsistencies due to limited local system knowledge. All agents compute their control inputs based on local subsystem knowledge while only predicting the control inputs of the neighboring agents. If a predicted control input diverts from the computed control input, this is called a prediction inconsistency and can lead to safety-critical failures. In this work, we show how to guarantee prediction consistency using synchronization based on \cite{kloock_coordinated_2023} and \cite{kloock_cooperative_2023}. Previous work on achieving prediction consistency in \ac{mycdmpc} includes sequential approaches and parallel approaches.  In sequential approaches, the agents compute sequentially and achieve prediction consistency by sharing their predicted outputs \cite{trodden_cooperative_2013,mueller_cooperative_2012,blasi_distributed_2018}.\\
    Achieving prediction consistency in parallel approaches is more difficult, but desirable as they are more scalable. The approaches explored in the literature can be clustered into different ideas. In \cite{trodden_adaptive_2009}, the coupling graph is divided into disjoint sub-graphs that are fully connected and can be analyzed independently. In \cite{maestre_distributed_2011}, the agents are given initial decisions and adapt these decisions slightly to reduce their local cost. This approach does not scale well, as it requires computing the initial decisions. The works of \cite{sahebjamnia_designing_2016,francisco_multi_2019,morales_influence_2019} use fuzzy logic to deal with the prediction inconsistencies. The authors of \cite{maxim_min_2019,fele_coalitional_2018,chanfreut_coalitional_2019} use coalitional games from game theory to cluster agents in the coupling topology, leading to potentially high node degrees. Instead of distributing the control problem, it is also possible to solve the centralized optimization problem using distributed optimization as shown in \cite{falsone_dual_2017,hashempour_distributed_2021,segovia_distributed_2021,hours_parametric_2016,braun_hierachical_2018}.
\section{Background}
\begin{figure}[!htb]
        \centering
        \begin{subfigure}{0.45\linewidth}
            \centering
            \scalebox{1}{
\begin{tikzpicture}[roundnode/.style={circle, draw, minimum size=7mm}, baseline=(current bounding box.center), node distance=1.5cm]
    \node[roundnode] (1) {$1$};
    \node[roundnode] (3) [below right of=1] {$3$};
    \node[roundnode] (4) [below left of=3] {$4$};
    \node[roundnode] (2) [below left of=1] {$2$};
    
    \draw[-] (1) edge[right=15] node [left] {} (3);
    \draw[-] (3) edge[right=15] node [left] {} (2);
    \draw[-] (4) edge[right=15] node [left] {} (2);
    \draw[-] (1) edge[right=15] node [left] {} (2);
\end{tikzpicture}
}
            \subcaption{Coupling graph $\graphUndirected$.}
            \label{subfig:coupling_graph}
        \end{subfigure}
        \begin{subfigure}{0.45\linewidth}
            \centering
            \scalebox{1}{
\begin{tikzpicture}[roundnode/.style={circle, draw, minimum size=7mm}, baseline=(current bounding box.center), node distance=1.5cm]
    \node[roundnode, draw=red!50, line width=1.5pt] (1) {$1$};
    \node[roundnode] (3) [below right of=1] {$3$};
    \node[roundnode, draw=white] (4) [below left of=3] {};
    \node[roundnode] (2) [below left of=1] {$2$};
    
    \draw[-] (1) edge[right=15] node [left] {} (3);
    \draw[-] (3) edge[right=15] node [left] {} (2);
    \draw[-] (1) edge[right=15] node [left] {} (2);
\end{tikzpicture}
}
            \subcaption{Coupling sub-graph $\graphUndirected_{1}$ for agent 1.}
            \label{subfig:coupling_subgraph}
        \end{subfigure}
        \caption{Example for (a) a coupling graph and (b) the corresponding coupling sub-graph for agent 1.}
        \label{fig:graphs}
    \end{figure}
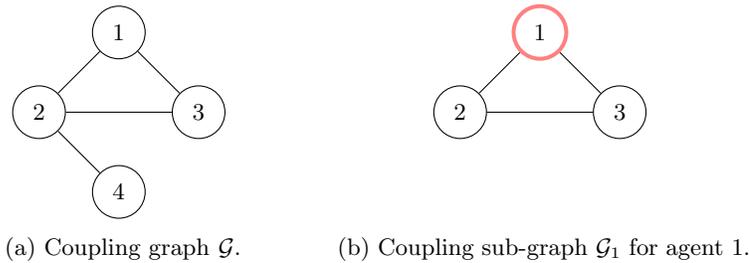

    We consider multiple agents in an \ac{ncs}. 
    The topology of the \ac{ncs} is modeled as a weighted undirected graph $\graphUndirected = (\setAgents,\setEdges,\setWeights)$ called coupling graph. 
    Here, $\setAgents$ represents the set of nodes, $\setEdges$ the set of edges and $\setWeights$ their corresponding weights. 
    Each node in the coupling graph represents an agent and the edges the coupling relations between them. 
    We define the coupling sub-graph $\graphUndirected_{i} = (\setAgents_{i}, \setEdges_{i},\setWeights_{i})$, consisting of agent $i$, its neighbors and the edges between them. Figure \ref{fig:graphs} shows an example coupling graph and the coupling sub-graph for agent 1.
    \\Consider the distributed control problem in an \ac{ncs}. At each time step, we aim to compute the optimal control inputs for each agent $i \in \setAgents$ to follow its reference trajectory $\trajectoryReference_i$. To solve this distributed control problem we use \ac{mycdmpc} as it allows us to incorporate individual as well as joint objectives and constraints.
    The set of agents that agent $i$ cooperates with is given by the set  $\setAgents_i$ in the coupling sub-graph $\graphUndirected_{i}$. We formulate the local \ac{mycdmpc} problem for agent $i$ as \\
        \begin{equation}
        \begin{aligned}
                &\begin{aligned}
                    \text{minimize}\quad & \begin{aligned}     
                    &\sum_{j\in \setAgents_{i}} \weight_{i\rightarrow j} \sum_{k=1}^{\horizonPrediction-1} \ell^x_{j}\left(\sysState_{i\rightarrow j}(k),\trajectoryReference_{j}(k)\right) +
                    \sum_{j\in \setAgents_{i}} \weight_{i\rightarrow j} \ell^f_{j}\left(\sysState_{i\rightarrow j}(\horizonPrediction) ,\trajectoryReference_{j}(\horizonPrediction)\right) +\\
                    &\sum_{j\in \setAgents_{i}} \weight_{i\rightarrow j} \sum_{k=0}^{\horizonControl-1} \ell_{j}^u\left(\Updelta \sysControlInputs_{i\rightarrow j}(k)\right) +
                    \sum_{(j, q)\in \setEdges_{i}} \sum_{k=1}^{\horizonPrediction}\ell^\text{c}_{j \rightarrow q}(\sysState_{i\rightarrow j}(k),\sysState_{i\rightarrow q}(k))\end{aligned}\\
                    \text{subject to}\quad &\forall j \in \setAgents_i, q \in \setAgents_i \cap \setAgents_j
                \end{aligned}\\
                &\begin{aligned}
                    \sysState_{i \rightarrow j}(k) &= f_{j}(\sysState_{i \rightarrow j}(k),\sysControlInputs_{i \rightarrow j}(k)), &\timestepIterator=1,\dots,\horizonPrediction-1,\\
                    \sysState_{i \rightarrow j}(k) &\in \setFeasibleStates_{j}, &\timestepIterator = 1,\dots, \horizonPrediction-1,\\
                    \sysState_{i \rightarrow j}(N_p)&\in \setFeasibleStates_{j}^\text{f},\\
                    \sysControlInputs_{i \rightarrow j}(k) &\in \setFeasibleInputs_{j}, &\timestepIterator = 0,\dots, \horizonControl-1,\\
                    \Updelta \sysControlInputs_{i \rightarrow j}(k) &\in \Updelta\setFeasibleInputs_{j}, &\timestepIterator=0,\dots,\horizonControl-1,\\
                    c^\text{c}_{j\rightarrow q}(\sysState_{i \rightarrow j}(k),\sysState_{i\rightarrow q}(k)) &\leq 0, &\timestep=1,\dots,\horizonPrediction,
                \end{aligned}
            \end{aligned}
            \label{eq:ocp_cdmpc}
        \end{equation}     
        where $\ell^x_{j}, \ell_{j}^\text{f}, \ell_{j}^u$ denote the reference deviation cost, the terminal cost, and the input variation cost for agent $j$, respectively, and $\ell_{j\rightarrow q}^\text{c}$ denotes the coupling objective between agent $j$ and agent $q$. 
        The variables $\sysState_{i\rightarrow j}(k)$, $\sysControlInputs_{i\rightarrow j}(k)$ and $\Updelta \sysControlInputs_{i\rightarrow j}(k)$ represent the state, the input, and the input variation at time step $k$ of agent $j$ predicted by agent $i$. The parameters $N_p$ and $N_u$ denote the prediction and the control horizon, respectively, and $\weight_{i\rightarrow j}$ represents the weight of the corresponding edge. 
        The function $f_{j}$ represents the system dynamics and $\setFeasibleStates_{j}, \setFeasibleStates_{j}^\text{f}, \setFeasibleInputs_{j}$, and $\Updelta\setFeasibleInputs_{j}$ represent the set of feasible states, the terminal set, the set of feasible inputs, and the set of feasible input variations, respectively. 
        The function $c^\text{c}_{j\rightarrow q}$ denotes the coupling constraint between agent $j$ and agent $q$.\\
        We outline the \ac{mycdmpc} procedure in Algorithm \ref{alg:agent-level_planning}.
            \begin{algorithm}[!b]
                \caption{\ac{mycdmpc} algorithm for agent $i$ \label{alg:agent-level_planning}}
                \begin{algorithmic}[1]
                \STATE \textbf{Input:} reference trajectories $\trajectoryReference_{j}$, $\forall j \in \setAgents_{i}$
                \STATE \textbf{Output:} control inputs $\sysControlInputs_{i\rightarrow j}$ and predicted states $\bm{x}_{i\rightarrow j}$
                \STATE Send and receive states to and from neighboring agents
                \STATE Solve \ac{mycdmpc} problem \eqref{eq:ocp_cdmpc} for $\sysControlInputs_{i\rightarrow j}$ and $\bm{x}_{i\rightarrow j}$ 
                \end{algorithmic}
            \end{algorithm}
        Each agent computes optimal inputs for itself and predicts the optimal inputs of its neighbors. Due to the limited local system knowledge of each agent, the predictions can be inconsistent between agents. 
    \section{The SCDMPC Algorithm}
        
        We propose to synchronize the states globally to guarantee prediction consistency. Our approach is inspired by multi-agent consensus \cite{ren_consensus_2005} and synchronization \cite{lunze_networked_2019}. To restore consistency, each agent $i$ synchronizes the states it predicted for its neighbors. Agent $i$ considers the states agent $j$ computed for itself, as well as the states the agents $q$ (that are connected to both agent $i$ and agent $j$) predicted for agent $j$. Specifically agent $i$ computes the weighted average  
        \begin{align}
            \bar{\bm{x}}_{i\rightarrow j} = \sum_{q \in \setAgents_{i} \cap \setAgents_{j}} \frac{1}{\weight_{q \rightarrow j}}\bm{x}_{q\rightarrow j}.
        \end{align}
        Similar to the agent-level planning, agent $i$ only considers a subset of the predictions of other agents that exist for agent $j$. Therefore, prediction consistency may not be satisfied after the first synchronization step. Consequently, we designed the synchronization as an iterative process. At every synchronization step, each agent synchronizes the states and communicates them to its neighbors. The size of these vectors scales linearly with the prediction horizon $N_\text{p}$ and the dimension of the state space, affecting the communication overhead as the number of agents or the complexity of the problem increases. However, as each agent only communicates with its neighbors, the overall communication remains distributed and scalable compared to centralized approaches. Then, each agent checks if the synchronized predictions from its neighbors are consistent with its own predictions. If the predictions are consistent, the synchronization procedure terminates, if not we repeat the procedure until we converged to a consistent solution. Theorem \ref{th:scdmpc-sync} states that the synchronization is guaranteed to converge and consequently terminate if the coupling sub-graph contains a spanning tree, i.e., if at least one agent has a path to all other agents.
        \begin{theorem} \label{th:scdmpc-sync}
            The synchronization converges to a solution if and only if each coupling sub-graph $\graphUndirected_{i}$ contains a spanning tree.
        \end{theorem}
        For the proof of Theorem \ref{th:scdmpc-sync}, see \cite{kloock_coordinated_2023}. We outline the system-level synchronization method in Algorithm \ref{alg:system-level_synchronization}. 
        \begin{algorithm}[!b]
            \caption{Synchronization algorithm for agent $i$ \label{alg:system-level_synchronization}}
            \begin{algorithmic}[1]
            \STATE \textbf{Input:} inconsistent states $\bm{x}_{q \rightarrow j}, \forall j \in \setAgents_i, q \in \setAgents_i \cap \setAgents_j$
            \STATE \textbf{Output:} consistent states $\bar{\bm{x}}_{j}$
            \STATE Initialize synchronized states $\bar{\bm{x}}_{q\rightarrow j} = \bm{x}_{q \rightarrow j}$
            \WHILE{predictions not consistent}
                    \STATE Send states $\bar{\bm{x}}_{i \rightarrow j}$ to agents $j$
                    \STATE Receive states $\bar{\bm{x}}_{q \rightarrow j}$ from agents $q$
                    \FORALL{$j \in \setAgents_i$}
                        \STATE $\bar{\bm{x}}_{i \rightarrow j} = \sum_{q} \frac{1}{\weight_{q \rightarrow j}}\bar{\bm{x}}_{q \rightarrow j}$
                    \ENDFOR
            \ENDWHILE
            \STATE $\bar{\bm{x}}_{j} =  \bar{\bm{x}}_{i \rightarrow j}$
            \end{algorithmic}
        \end{algorithm}
        Our complete \ac{myscdmpc} scheme is defined in Algorithm \ref{alg:scdmpc}. The algorithm loops until the resulting solutions are locally feasible and prediction consistent. Within this loop, each agent first computes a solution for itself and its neighbors using the \ac{mycdmpc} procedure given in Algorithm \ref{alg:agent-level_planning}. Then the corresponding predictions are synchronized using the synchronization procedure given in Algorithm \ref{alg:system-level_synchronization}. If the resulting predictions are feasible, the algorithm terminates; if not, the process is repeated using the synchronized predictions as the reference in the next iteration of the loop. If the solution space of the optimization problem is convex, \ac{myscdmpc} terminates after the first iteration. For a non-convex solution space we state that there exists a coupling topology such that \ac{myscdmpc} is guaranteed to converge to a feasible solution, if a feasible solution to the corresponding centralized optimization problem exists. For a proof see \cite{kloock_coordinated_2023}. Therefore, the convergence and consequently the termination of the algorithm depends on the coupling topology and the corresponding coupling weights.
        \begin{algorithm}[!htb]
        \caption{\acl{scdmpc} for agent $i$ \label{alg:scdmpc}}
        \begin{algorithmic}[1]
            \STATE \textbf{Input:} reference trajectories $\trajectoryReference_j, \forall j \in \setAgents_i$, indices $j_\text{s} \in \setAgents_i, q \in \setAgents_i \cap \setAgents_j$
            \STATE \textbf{Output:} control inputs $\sysControlInputs_{i\rightarrow j}$ and predicted states $\bm{x}_{i\rightarrow j}$
            \WHILE{states are not feasible and prediction consistent}
                \STATE $(\sysControlInputs_{i \rightarrow j},\bm{x}_{i \rightarrow j}) \gets$ CDMPC$(\trajectoryReference_j)$ (Alg. \ref{alg:agent-level_planning})
                \STATE Receive predictions $\bm{x}_{q\rightarrow j}$
                \IF{predictions inconsistent}
                    \STATE $\bm{x}_{j} \gets \text{Synchronization}(\bm{x}_{q \rightarrow j})$ (Alg. \ref{alg:system-level_synchronization})
                \ENDIF
                \STATE $\trajectoryReference_{j} \gets \bm{x}_{j}$
            
            \ENDWHILE
            \end{algorithmic}
        \end{algorithm}
        \vspace{-1cm}
    \section{Evaluation}
\begin{wrapfigure}{r}{0.515\textwidth}
            \centering
            \includegraphics[width=1\linewidth]{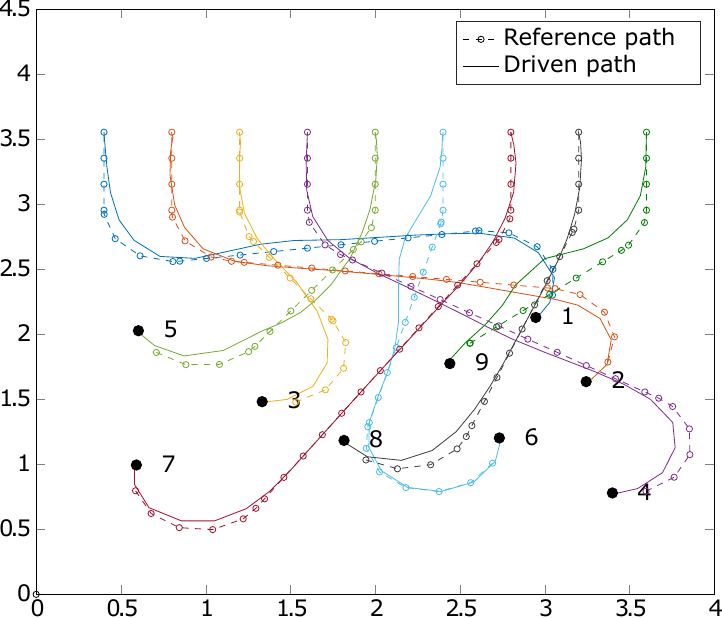}
            \caption{Visualization of the formation building scenario. }
            \label{fig:g2f_paths}
        \end{wrapfigure}

\begin{figure}[!b]
        \centering
        \begin{subfigure}{0.45\linewidth}
            \centering
            \includegraphics[width=\linewidth]{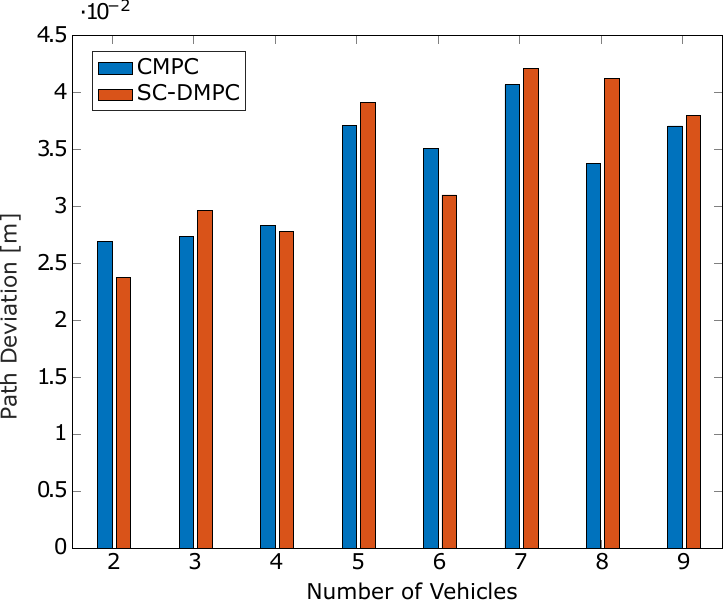}
            \subcaption{Path deviation.}
            \label{subfig:path_dev}
        \end{subfigure}
        \hspace{0.5cm}
        \begin{subfigure}{0.48\linewidth}
            \centering
            \includegraphics[width=\linewidth]{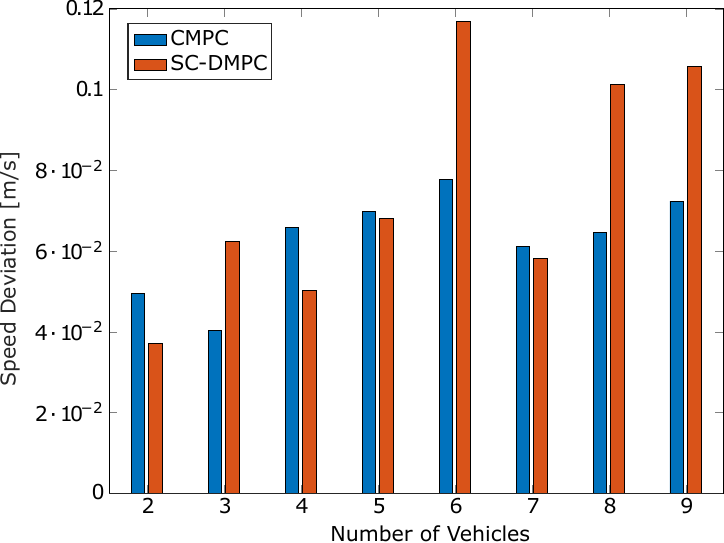}
            \subcaption{Speed deviation.}
            \label{subfig:speed_dev}
        \end{subfigure}
        \caption{The average cumulative path and speed deviations of \ac{mycmpc} and \ac{myscdmpc} for different numbers of \acp{cav}.}
        \label{fig:cumulative_deviations}
    \end{figure}

        We evaluate the \ac{myscdmpc} algorithm in our CPM Lab \cite{kloock_cyber_2021}, an open-source, remotely accessible small-scale test-bed using \acp{cav} in 1:18 scale. Each vehicle executes the algorithm on a designated computation node, a real-time Ubuntu 18.04 with two \SI{1.6}{\giga\hertz} cores and \SI{16}{\giga\byte} of RAM. Each vehicle communicates with its computation node via WiFi. The computation nodes share the vehicle's information via Ethernet. The algorithm was implemented in MATLAB R2020a. To solve optimization problem \eqref{eq:ocp_cdmpc}, we use the optimization toolbox of IBM CPLEX 12.10. For the costs we use the weighted $\ell_2$-norm while omitting the coupling objective. The kinematic bicycle model predicts system dynamics, while coupling constraints enforce safety distances between agents.\\ 
        We test the algorithm in a formation building scenario. A visualization of the scenario is given in Figure \ref{fig:g2f_paths}. Each vehicle is placed at a random location on the \SI{4}{\meter} x \SI{4}{\meter} driving area of the test-bed. The driving area itself does not contain obstacles and is therefore convex. The \acp{cav} aim to form a predefined formation, in our case standing next to each other at the top of the driving area.
        The controller of each vehicle computes the optimal trajectory to arrive at the goal pose as fast as possible while avoiding collisions. We generate a reference path using Dubins Path algorithm and compute a respective reference trajectory by sampling points on the reference path based on the speed of the vehicle. 
        \setlength{\intextsep}{0pt}%
        For more information we refer to \cite{kloock_cooperative_2023}. We run the scenario multiple times with random start poses and fixed goal poses using \ac{mycmpc} and \ac{myscdmpc} and compare the results.\\
        The average cumulative path and speed deviations shown in Figure \ref{fig:cumulative_deviations} indicate that the deviations are similar for \ac{myscdmpc}  and \ac{mycmpc}. In some instances \ac{myscdmpc} achieves smaller deviations than \ac{mycmpc}. 
    \begin{figure}[!t]
        \centering
        \includegraphics[width=0.5\linewidth]{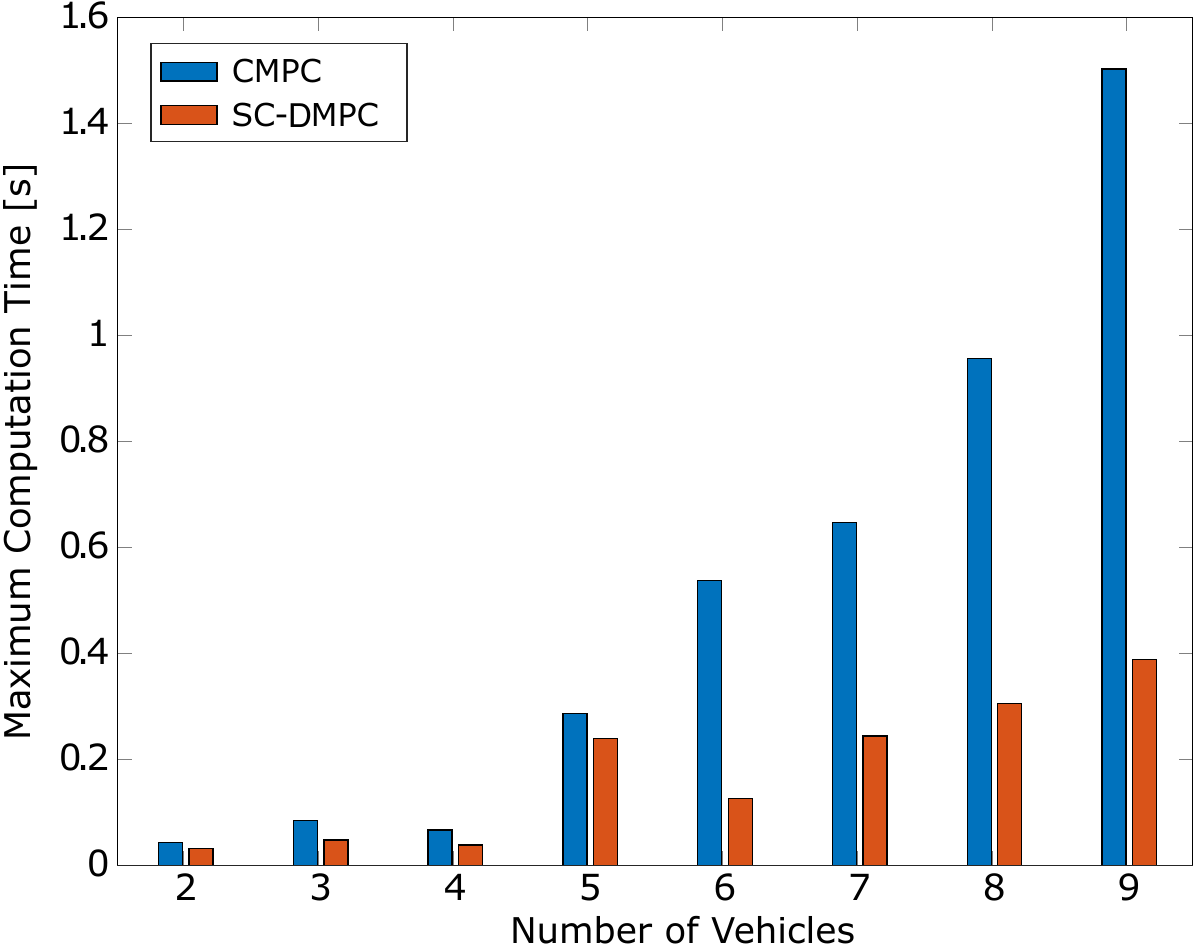}
        \caption{The maximum computation time of \ac{mycmpc} and \ac{myscdmpc} for different numbers of \acp{cav}.}
        \label{fig:comp_time}
    \end{figure}
    In those cases \ac{myscdmpc} had a higher cost for the control input variations or the coupling objective than \ac{mycmpc}, which are not considered here. Furthermore, it can be observed that the average cumulative path deviations increase with the number of \acp{cav}. In Figure \ref{fig:comp_time}, the maximum computation time of \ac{myscdmpc} and \ac{mycmpc} is shown for different numbers of \acp{cav}. It is evident that, for both the maximum computation time increases with the number of \acp{cav}. However, the \ac{myscdmpc} approach proves to be more scalable as the maximum computation time increases at a slower rate. Note, that the computation time of Algorithm \ref{alg:scdmpc} depends on the coupling topology of the agent. The number of neighbors of an agent increases the number of optimization variables and constraints and consequently the computation time needed to solve the optimization problem.

    \section{Conclusion}
    This work presented a Synchronization-Based Cooperative Distributed Model Predictive Control approach. We show that through synchronization, prediction consistency can be guaranteed. In our experiments in the CPM Lab we demonstrate the applicability of the \ac{myscdmpc} algorithm for planning trajectories of \acp{cav}. The results make it evident that the \ac{myscdmpc} achieves control performance close to that of \ac{mycmpc} while showcasing better scalability in terms of computation time. In future research, we will investigate the effect of communication delays as well as learning-based approaches to enhance the performance and safety of our approach.
%
%

\clearpage

\end{document}